# Boosting Perovskite Solar Cell Stability: Dual Protection with Ultrathin Plasma Polymer Passivation Layers

Mahmoud Nabil[a], Lidia Contreras-Bernal*[b,c], Gloria P. Moreno-Martinez[b], Jose Obrero-Perez[b], Javier Castillo-Seoane[b], Juan A. Anta[a], Gerko Oskam[a], Paul Pistor[a], Ana Borrás[b], Juan R. Sanchez-Valencia[b], Angel Barranco*[b]

[a]Center for Nanoscience and Sustainable Technologies (CNATS). Department of Physical, Chemical, and Natural Systems. Universidad Pablo de Olavide. 41013 Sevilla, Spain
[b]Institute of Materials Science of Seville, Spanish National Research Council. Universidad de Sevilla, C. Américo Vespucio 49, Seville, Spain
[c]Dpto. Ingeniería y Ciencia de los Materiales y del Transporte. Escuela Politécnica Superior Universidad de Sevilla. c/ Virgen de África 7, Seville E-41011, Spain
lcbernal@us.es; angel.barranco@csic.es

## Abstract

Metal halide perovskite solar cells (MHPSCs) hold great promise related to their high efficiency and low fabrication costs, but their long-term stability under environmental conditions remains a major challenge. In this study, we demonstrate an effective protection strategy to enhance the stability of MHPSCs through the incorporation of a double passivation layer based on an adamantane-based plasma polymer (ADA) at both the electron transport layer (ETL)/perovskite and perovskite/hole transport layer (HTL) interfaces. Our results show that the implemented ADA deposition technique is compatible with delicate substrates such as perovskites thin films, as their optical, morphological and optoelectronic properties are unaltered upon ADA deposition. At the same time, it provides effective protection to the perovskite material in high humidity environments. The ADA-double passivation not only reduces the formation of mobile ionic defects that cause additional recombination, but also significantly reduces humidity-induced degradation and mitigates the photocatalytic degradation caused by $TiO_2$ under UV exposure. Stability tests performed under 100% relative humidity and continuous AM 1.5G illumination (ISOS-L-1) show that ADA dual passivated devices retained 80% of their initial efficiency after 4000 minutes, while reference samples dropped to 30%. The enhanced performance of ADA-passivated cells is attributed to protective nature of the plasma polymer layer resulting in a preservation of photocurrent and the prevention of new recombination routes. This dual passivation strategy offers a promising route to improve the environmental stability of PSCs and extend their operational lifetime.

## Introduction

Third generation solar cells based on metal halide perovskite (MHPSCs) have proven to be a competitive technology in terms of efficiency and manufacturing cost.[1–3] However, these devices show limited long-term environmental stability. Continuous illumination, high temperatures (above 60 ºC), oxygen, and moisture significantly affect to the photovoltaic (PV) performance of MHPSCs.[4–6] The stability of perovskite solar cells depends both on the nature of the perovskite material itself and on the device architecture. On the one hand, the perovskite material with PV properties consists of an $ABX_3$ type structure (where A is a monovalent cation, B a divalent cation, and X are halide ions) with a hygroscopic character, and partially volatile species, making it unstable in presence of environmental stimuli.[7–10] Perovskite materials exhibit a relatively high density of defects and a highly dynamic crystal network, which leads to ionic migration and, ultimately, decomposition.[11] Additionally, they may undergo phase segregation when exposed to light.[12] All these phenomena affect the MHPSCs stability.[4] On the other hand, the conventional structure of a highly efficient MHPSC (*n-i-p* configuration) requires that the perovskite layer be sandwiched between an electron selective layer (ETL) and a hole selective layer (HTL) to extract the charge photogenerated within the perovskite material. In general, top metal electrodes are required to close the electrical circuit.[13] The physical and chemical characteristics of these selective layers significantly affect the stability of the devices. For instance, one of the most popular and efficient single-junction MHPSCs is based on a mesoporous scaffold of $TiO_2$ as the ETL and spiro-OMeTAD layer as the HTL. The photocatalytic behavior of $TiO_2$, when it is exposed to UV radiation, triggers a decomposition reaction in the perovskite.[14] Moreover, the $TiO_2$ scaffold is often doped with hygroscopic materials to enhance the electron mobility, which induces instability of the solar devices under humid ambient conditions.[15,16] Spiro-OMeTAD requires dopants to improve its hole mobility, such as TBP or Li salts, which corrode or diffuse into the perovskite, accelerating its degradation.[17,18] In addition, metal species from electrodes can migrate through the HTL and contribute to perovskite degradation, while halides from the perovskite can corrode the metal contacts, reducing the conductivity over time.[19] It is noteworthy that a high density of surface defects in the perovskite further accelerates degradation while ion migration and phase segregation within the perovskite layer is also an important source of instability and degradation.[20–22] Similar

effects are observed when PSCs suffer from interfacial recombination and charge extraction losses.[23,24]

In this context, several strategies to improve MHPSCs stability haven been explored in recent years, including: modify perovskite composition to reduce ionic mobility and increase crystalline stability, or even enhancing the hydrophobicity;[25–27] protecting the perovskite layer and selective contacts from external factors via encapsulation;[28–30] developing robust selective contacts and relevant additives for stable PSCs;[17] and passivating interlayers to act as a barrier against environmental factors and reduce the ion migration and accumulation at interfaces.[4,23,31–35]

Focusing on interface engineering through the incorporating passivation layers at interfaces, in PSCs based on a *n-i-p* architecture, the primary focus has been on *in situ* passivation of the ETL prior to the perovskite deposition, as the ETL is generally more robust than the perovskite material and thus withstands surface treatments better. Of the few studies that focus on the perovskite/HTL interface, these often involve solution methods, such as the incorporation of polymers (PVDF, PTAA and PTPD) in the antisolvent step, ammonium salts, or the formation of 2D perovskites, among others.[36–39] No alternative vacuum processes scalable to industrial manufacturing have been explored.

Herein, we propose the use of an ultrathin adamantane plasma polymer film (ADA) deposited by remote plasma-assisted vacuum deposition (RPAVD) technique as passivation material for both the ETL/perovskite and the perovskite/HTL in a *n-i-p* configuration of perovskite solar cells. The plasma polymer film is developed in a single step through the vacuum sublimation of an adamantane powder precursor in presence of a remote microwave plasma (see Scheme 1). RPAVD is carried out at room temperature and is compatible with highly sensitive substrates and large areas.[40,41] The resulting ADA plasma polymer film is transparent in the visible and UV ranges, thermally stable up to 250 ºC, insoluble in water, and serves as a conformal protective layer for organic supported nanostructures.[42] Thick layers of ADA have also been shown to function as effective encapsulants for PSCs, protecting them from water immersion and humidity.[28] Furthermore, in a recent publication, we demonstrated that the ADA film is a reliable passivation layer for the ETL/perovskite interface, improving both moisture stability and device reproducibility without compromising the efficiency of the PSCs.[34] It is important to note that using the same material for different cell functions has the singular benefit of facilitating the large-scale production of these devices.

Thus, we first verified the compatibility of the RPAVD technique with sensitive materials such as organometallic perovskites using UV and FTIR spectroscopies, as well as X-ray photoelectron spectroscopy (XPS). Once the optimum thickness of the ADA film for the perovskite/HTL interlayer was determined by ensuring it did not affect the PV performance of the device, the environmental stability of the devices was analyzed by measuring and comparing the photovoltaic performance and the impedance response of fresh and aged devices. Finally, solar cells incorporating double passivation layers, sandwiching the perovskite material between ADA layers, were fabricated. The long-term stability of the devices was evaluated under 100% relative humidity following the ISO-L-1 procedure.[43]

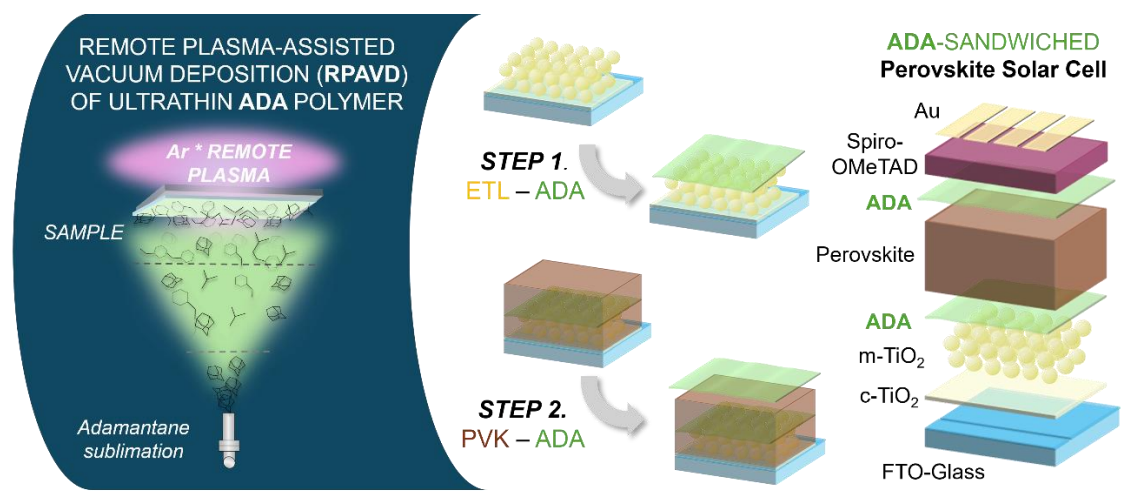

**Scheme 1.** Schematic of the solar cell preparation process for ADA-based sandwiched architecture.

## Materials and Methods

### Materials

To fabricate the perovskite solar cells based on RbCsMAFA perovskite [(FAPbI$_3$)$_{83}$(MAPbBr$_3$)$_{17}$ + 5% CsI) + 5% RbI], we utilized a variety of high-purity materials and solvents to ensure the quality and reproducibility of the devices. Fluorine-doped tin oxide (FTO) coated glass substrates (TEC 15) with a sheet resistance of 12-15 Ω/sq and a transmittance of 82-84.5% were obtained from Pilkington; these substrates were pre-etched to separate charge collection areas. Titanium diisopropoxide bis(acetylacetonate) (75% in 2-propanol) and TiO$_2$ paste (18NRT) were sourced from Sigma-Aldrich, while absolute ethanol (99.9%) used for dilution was procured from Scharlau. Solvents including N,N-dimethylformamide (DMF), dimethyl sulfoxide (DMSO), chlorobenzene (ClBn), and acetonitrile (AN) were supplied by Acrós Organics. Perovskite precursors such as formamidine hydroiodide (FAI, >98%), methylamine hydroiodide (MAI, >99.0%), lead (II) iodide (PbI$_2$, 99.99%), and lead (II) bromide (PbBr$_2$, 99.99%) were purchased from TCI. Cesium iodide (CsI, 99.9%) was obtained from Alfa Aesar, and rubidium iodide (RbI, 99.9%) was sourced from Sigma-Aldrich. For the hole transport material, 2,2,7,7-tetrakis[N,N-di(4-methoxyphenyl)amino]-9,9-spirobifluorene (spiro-OMeTAD, >99.9%) was purchased under the name SHT-263 Solarpur from Sigma-Aldrich. Additionally, lithium bis(trifluoromethanesulfonyl)imide (LiTFSI, 99.95%), tris(2-(1H-pyrazol-1-yl)-4-tert-butylpyridine)-cobalt(III)tris(bis(trifluoromethylsulfonyl)imide) (FK209 Co(III)), 98%), and 4-tert-butylpyridine (98%) were also procured from Sigma-Aldrich. Au 1-3 mm pellets (99.99% pure) were purchased from Testbourne Ltd. Adamantane powder (≥99%), from Sigma-Aldrich, was used for the remote plasma-assisted vapor deposition of interlayers. All materials were used as received without further purification to maintain consistency and accuracy in the experimental procedures.

**Device fabrication**

For the reference devices, the fluorine-doped tin oxide (FTO) substrates were first brushed with a solution of Hellmanex III© in water (2:98 vol %) and rinsed with deionized water. The substrates then underwent sequential ultrasonic cleaning of 15 minutes each in Hellmanex III© solution, deionized water, acetone, and isopropanol. After cleaning, the substrates were dried with nitrogen gas and treated with UV/O$_3$ for 15 minutes using an Ossila© UV/Ozone Cleaner. A 30 nm compact TiO$_2$ (c-TiO$_2$) layer was subsequently deposited onto the substrates via the spray pyrolysis method. A precursor solution was prepared by mixing 1 mL of titanium diisopropoxide bis(acetylacetonate) (75% in 2-propanol, Sigma Aldrich©) with 14 mL of absolute ethanol. This solution was immediately sprayed onto the pre-heated substrates at 450 °C for 30 minutes, using oxygen as the carrier gas. The substrates were masked before to avoid bottom electrode coverage. After deposition, the substrates were allowed to cool down to room temperature, to then being UV/O$_3$ treated for 15 minutes. A mesoporous TiO$_2$ (m-TiO$_2$) solution was prepared by adding 1 mL of absolute ethanol to 150 mg of commercial TiO$_2$ paste (18NRT, Sigma Aldrich©), and the mixture was stirred overnight. This solution was spin-coated onto the c-TiO$_2$ layer at 4000 rpm for 10 seconds. The films were then heated on a hot plate set at 120°C, followed by ramping up to 450 °C and maintaining this temperature for 30 minutes before cooling.

The FTO/c-TiO$_2$/m-TiO$_2$ substrates underwent a UV/O$_3$ treatment for 15 minutes before being transferred to a nitrogen glovebox with O$_2$ and H$_2$O levels below 0.5 ppm and a temperature range of 25-28 °C for the perovskite layer deposition. For the perovskite precursor, two initial solutions were prepared: PbI$_2$ and PbBr$_2$, both at 1.5 M concentrations in a 1:4 volume ratio of DMSO to DMF. These solutions were heated to 60 °C until fully dissolved, then left to cool. Subsequently, formamidine hydroiodide (FAI) and methylammonium bromide (MABr) were added to the PbI$_2$ and PbBr$_2$ solutions, respectively, to form FAPbI$_3$ and MAPbBr$_3$ upon dissolution. These two solutions were then mixed in a 5:1 Vol % ratio of FAPbI$_3$ to MAPbBr$_3$. Finally, 5 Vol % of 1.7 M CsI solution in DMSO and 5 Vol % of RbI solution in a 1:4 Vol % DMSO: DMF were added to the mixture. The perovskite film was deposited using a two-step spin-coating process: the first step at 1000 rpm for 10 seconds, followed by 6000 rpm for 20 seconds. During the second step, 200 μL of chlorobenzene was added as an antisolvent 15 seconds after the beginning. The samples were then annealed at 100 °C for 60 minutes.

The hole transport layer was deposited by spin-coating a solution of 0.07 M spiro-OMeTAD in chlorobenzene, doped with LiTFSI (1.8 M in acetonitrile), FK209 Co (III) (0.25 M in acetonitrile), and 4-tert-butylpyridine in a molar ratio of 0.5, 0.03, and 3.3, respectively. This solution was filtered before being spin-coated in two steps: 200 rpm for 10 seconds followed by 6000 rpm for 10 seconds. The substrates were then taken out of the glovebox, and a small portion of the deposited m-TiO$_2$/perovskite/spiro-OMeTAD layers was selectively removed to expose the underlying FTO layer. This exposed area overlapped with the part covered during the compact layer deposition. Finally, approximately 70-80 nm gold electrodes were deposited by thermal evaporation at a pressure below $10^{-6}$ Torr. The assembled solar cells were then left to stabilize overnight in a dry box under ambient conditions.

**Integration of adamantane plasma polymer layer on the perovskite solar cells**

The deposition of the adamantane plasma polymer (ADA) thin films on top of perovskite layer (TiO$_2$/PVK/ADA) were performed in a microwave plasma reactor under a base pressure of approximately $10^{-6}$ torr, following methodologies described in our previous studies.[28,34,42,44,45] An electron cyclotron resonance (ECR) plasma reactor was utilized, featuring distinct zones for plasma generation and remote deposition. It is named as "plasma region" because the discharge is confined by the magnetic field provided by a set of magnets. An argon microwave plasma of 2.45 GHz was maintained at a power setting of 210 W and a pressure of $10^{-2}$ torr. The substrates were positioned 9.5 cm downstream from the plasma discharge area, ensuring they were facing the opposite direction of the plasma to ensure the remote condition. Adamantane powder was sublimated into the chamber during the deposition process by heatable dispenser outside the chamber, due to the high vapour pressure at room temperature. The thickness of the films and the deposition rate were continuously monitored using a quartz crystal microbalance (QCM) placed at the level of the substrate holder. A flat silicon wafer substrate is used for post-deposition thickness monitoring.

To prepare the perovskite sandwiched architecture between two ADA passivating layers (TiO$_2$/ADA/PVK/ADA/Spiro), additionally, 5 nm of ADA layer was deposited on top of the m-TiO$_2$ layer prior perovskite deposition, using the same methodology described above. In this case, ADA layer was treated with UV/O$_3$ for 5 minutes before being transferred to the glovebox for perovskite layer deposition.[34]

**Characterization of films and devices**

To determine the optical properties and thickness of the deposited ultrathin plasma films, variable angle spectroscopic ellipsometry (VASE) measurements were conducted using a Woollam© V-VASE ellipsometer. The optical constants, such as the refractive index and the film thickness, were obtained by fitting the ellipsometry spectra to the Cauchy model. This method allowed for precise thickness monitoring of the ultrathin ADA interlayers deposited on flat silicon wafer substrates. The optical absorption properties of the films were studied using a PerkinElmer© Lambda 750 UV/vis/NIR spectrophotometer, covering the wavelength range of 400-800 nm, which encompasses the typical absorbance range of perovskite materials. This technique was employed to compare the absorbance spectra of the perovskite films with and without the ultrathin ADA layer on top. Both systems were encapsulated with poly(methyl methacrylate) (PMMA, Sigma Aldrich©) to prevent from any potential degradation effect. The wetting properties of the films were characterized by measuring the static water contact angle (WCA) using a DataPhysics OCA 20 goniometer. A 1 µL droplet of deionized water was dropped on the film surface. The measurements were repeated two times for verification, ensuring accurate and reliable results. Stationary photoluminescence (PL) and time-resolved photoluminescence (TRPL) measurements were performed using an FLS1000 Photoluminescence Spectrometer (Edinburgh Instruments). For the stationary PL measurements, a 450-watt Xenon lamp was used as the excitation source at 465 nm. The emission scan ranged from 650 to 900 nm with a step size of 1 nm and a dwell time of 0.5 s. These measurements were performed on $TiO_2$/PVK and $TiO_2$/PVK/ADA films deposited on fused silica substrates. Fourier transform infrared spectroscopy (FTIR) spectra were obtained using a JASCO© FT/IR-6200 IRT-5000 spectrometer under vacuum conditions and transmittance mode for the layer that were deposited on intrinsic silicon wafers. Scanning electron microscopy (SEM) micrographs were obtained using a Hitachi© S4800 microscope operated at 2 kV. X-ray diffractograms were recorded on a Rigaku diffractometer using a CuKα source. The measurements were conducted in grazing angle geometry to enhance surface sensitivity. The samples were mounted without any modification, and the divergence slit was adjusted to match the dimensions of the films, ensuring optimal beam alignment. A scan range of 10–60° was chosen to encompass key diffraction peaks, with a scanning rate of 3° $min^{-1}$ to balance resolution and data acquisition time. Baseline correction was applied to the diffractograms to mitigate noise contributions from the substrate.

Current-voltage (*J-V*) curves of the fabricated devices were recorded under a solar simulator (ABET©-Sun2000) equipped with an AM 1.5 G filter at 100 mW/cm². The measurements were performed in reverse scan with a scan rate of 100 mV/s from 1.2 to -0.1 V. The samples were measured with a black mask leaving an active area of 0.14 cm². A statistical analysis of variance (ANOVA) was conducted on 12 devices of the structure analyzed. Electrochemical impedance spectroscopy (EIS) measurements were performed using the PAIOS characterization unit (Fluxim inc.). The measurements were conducted across a wide range of DC light intensities under white, blue (455 nm), and red (656 nm) illumination. For each measurement, the offset voltage was adjusted to the open-circuit photovoltage, and a 15 mV perturbation amplitude was applied in the frequency range of $10^6$ MHz to 1 Hz. After the initial measurements, the devices were stored in the dark under ambient conditions at approximately 50% relative humidity and ~25 °C for 48 days (>1100 hours). EIS measurements were then repeated on the same devices to study the degradation dynamics. To extract the EIS parameters, the Z-view modeling software (Scribner) was used to fit the spectra to an equivalent circuit of the form -$R_s$-($R_{LF}C_{LF}$)-($R_{HF}C_{HF}$)-,[46] where circuit elements in brackets represent parallel RC (resistor-capacitor) elements. Following these procedures, low and high frequency resistance and capacitance were extracted for all samples as a function of the open-circuit voltage.

To demonstrate the passivation effect of ADA layer on the photocatalytic properties of $TiO_2$ as initiator of the perovskite degradation, we carried out a photo-stability test consisting of the constant illumination of the samples with UV light to ensure the $TiO_2$ activation ($\lambda$ = 325 nm, P = 8.0 mW·cm$^{-2}$) and the acquisition of the photoluminescence of the perovskite material in real time. We used fused silica substrates with a c-$TiO_2$ and m-$TiO_2$ layers equivalent to PV devices. The illumination was incident on the back of the samples through the $TiO_2$ layers during 20 h. Steady-state spectra were measured before and after stability tests. The photo-stability was tested using a Jobin Yvon Fluorolog-3 spectrofluorometer from HORIBA© equipped with a 450 W continuous-wave xenon arc lamp as the excitation source. To prevent environmental degradation, the samples were encapsulated in a polymethyl methacrylate (PMMA) polymeric matrix. PMMA was dissolved at 5 wt.% in chlorobenzene. The solution was spin-coated on top of samples in two steps: 2000 rpm during 10 s (step 1) and 5000 rpm during 30 s (step 2).

Customized stability characterization tests were performed in different environments to evaluate the device behavior under extreme conditions. These tests employed a specially designed automated system that sequentially monitors full *J-V* curves and conducts statistical analysis of cell performance. An ABA LED Solar Simulator (Newport Model LSH-7320), calibrated for 1 sun illumination, was used to provide the light source. The samples were placed inside a software-controlled, homemade hermetic container capable of measuring four cells (16 electrodes) simultaneously. To maintain a controlled atmosphere, a continuous gas flow of 120 sccm was regulated by a rotameter to provide a dry atmosphere of nitrogen ($N_2$), oxygen ($O_2$), or air ($O_2/N_2$) within the container. When humidity is needed, the gas is passed through a water bubbler to adjust the relative humidity. A humidity sensor at the container outlet recorded the actual moisture levels. The devices were masked to expose an active area of 0.14 cm² for accurate photovoltaic parameter measurements. The stability tests were carried out using a fixed voltage bias near the maximum power point (MPP). FTO/c-$TiO_2$/m-$TiO_2$/PVK/ADA/Spiro/Au devices were tested under direct airflow and then under 80% humidity ($N_2$ as the carrier gas). FTO/c-$TiO_2$/ADA/m-$TiO_2$/PVK/ADA/Spiro/Au devices were tested under 100% humidity ($N_2$ as the carrier gas) at room temperature. Both were compared to reference cells and following the ISOS-L-1 protocol.[43]

## Results and Discussion

### Adamantane Plasma Polymers at perovskite/HTL interface

**Figure 1** and Table S1 compare the current density-voltage curves (*J-V* curves) and photovoltaic (PV) parameters of perovskite solar cells incorporating an ultrathin adamantane-based plasma polymers (ADA) layer at perovskite/hole transport layer (HTL) interface (see **scheme 1**), measured under 1 sun- AM 1.5G illumination. Three different ADA layer thicknesses - 30, 20 and 6 nm - were explored as passivating films for this interface; the ADA film thickness was determined by variable angle spectroscopic ellipsometry (VASE) (see Table S2). The solar cell configuration utilized was an *n-i-p* mesoporous structure with $TiO_2$ as the electron transport layer (ETL) and spiro-OMeTAD as HTL. ADA plasma polymer films were deposited onto the RbCsMAFA perovskite layer via the remote plasma-assisted vacuum deposition (RPAVD) technique, as previously reported. [28,34] No additional treatment of the polymeric layer was required before incorporating the HTL. Statistical analysis reveals that the thicker ADA layers adversely affect the PV performance of solar devices. In particular, the 30 nm ADA layer significantly reduces the fill factor (*FF*), open-circuit potential ($V_{OC}$), and short-circuit current density ($J_{SC}$) of resulting perovskite solar cell. Similarly, the 20 nm ADA layer mainly decreases the $J_{SC}$ compared to the reference samples.

In contrast, the presence of a 6 nm ADA layer at the perovskite /Spiro interface does not significantly alter the PV parameters relative to the reference values (see Table S1). In fact, the 6 nm ADA layer improves the reproducibility of PV parameters compared to reference samples. This trend (where thinner ADA layer result in better photovoltaic efficiency) is consistent with the understanding that the ADA polymeric film is a dielectric material which hinders the flow of electric current once its thickness surpasses a certain threshold.[34,42]

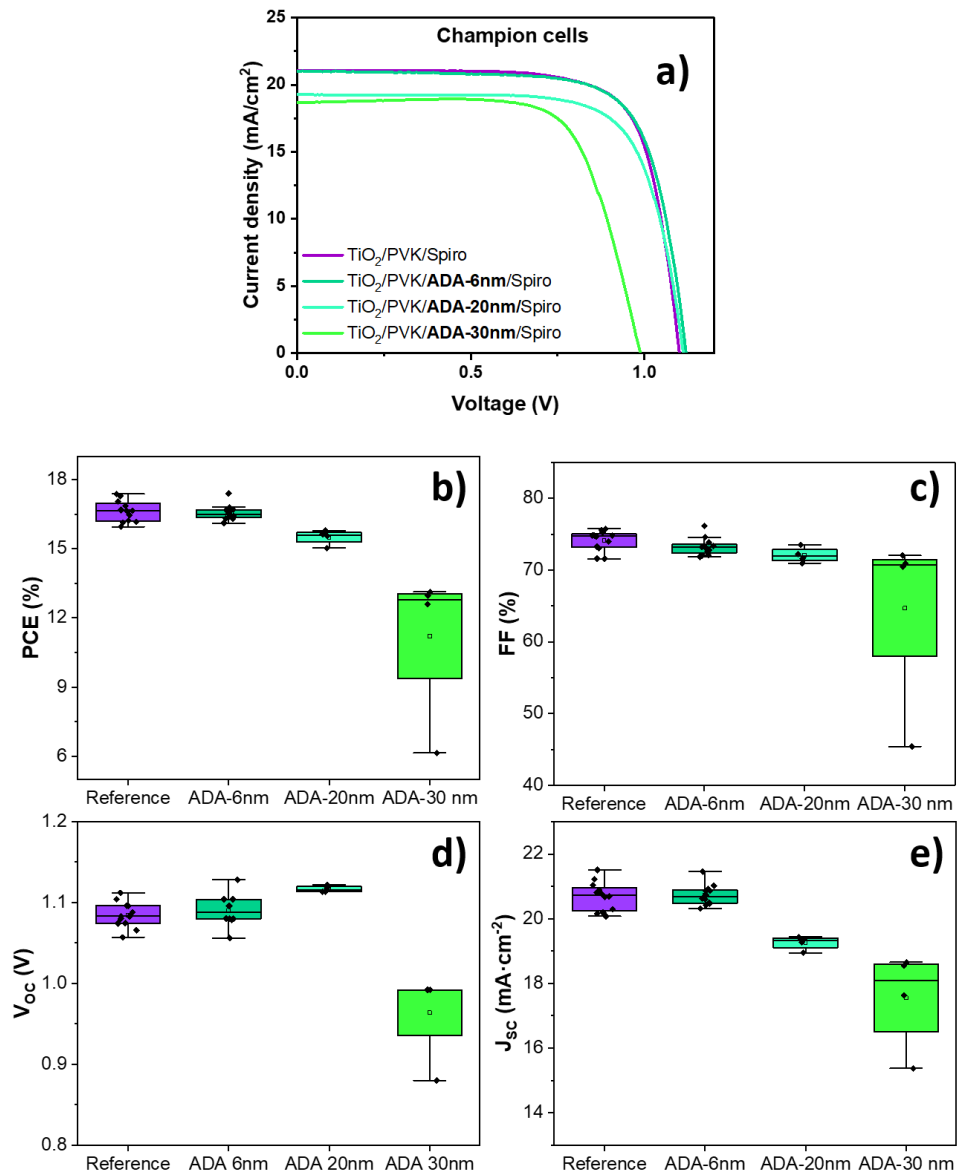

**Figure 1**. a) Current-density curves of perovskite solar cells containing adamantane plasma polymer films as a passivating layer at perovskite/HTL interface. b-e) Photovoltaic parameters statistics using analysis of variance. These data have been obtained at 100 mV·s$^{-1}$ reverse scan and using a mask of 0.14 cm$^2$.

Focusing on the optimal ADA thickness, Figure S1 shows the cross-sectional image of a complete PSC containing the passivating layer at the perovskite/HTL interface. It is observed that the 6 nm passivating layer does not induce any discernible alteration in the perovskite grain structure, nor does it affect the thickness of either the active layer (~700 nm) or the HTL (~200 nm) when compared to the reference samples.

**Figure 2**a illustrates the UV-vis absorption spectra of the perovskite material with and without the ADA film, deposited on mesoporous TiO$_2$ (hereafter referred to as TiO$_2$/PVK/ADA sample and TiO$_2$/PVK sample, respectively). The UV-Vis spectra are nearly identical and display similar absorption band intensities in the 750 nm region for both configurations, coinciding with the comparable thickness of the perovskite layer. This indicates that the incorporation of ADA does not cause any optical losses. Figure 2a also shows the photoluminescence (PL) spectra of both types of samples. The data have been normalized to the maximum absorbance for each device to enable a clearer comparison of the perovskite films. A somewhat higher PL peak intensity is observed for the TiO$_2$/PVK/ADA sample. This indicates that the ADA film does not accelerate non-radiative recombination and may even passivate the interface.

The optical analysis was further completed by water contact angle (WCA) measurements, as shown in Figure S2, which provides insights into the wettability properties of the surface. The WCA values for the 6 nm ADA layer and the perovskite layers in the TiO$_2$/perovskite/ADA and TiO$_2$/perovskite samples were 82° and 67°, respectively. The WCA observed for the ADA-passivated electrode corresponds to a partially hydrophobic surface and aligns with the value reported for a 200 nm ADA layer in our report,[28] despite of the significantly reduced thickness used in this work. Additionally, it is noteworthy that the water droplets deposited into both electrodes (with and without ADA layer), turned yellow indicating the presence of dissolved PbI$_2$.

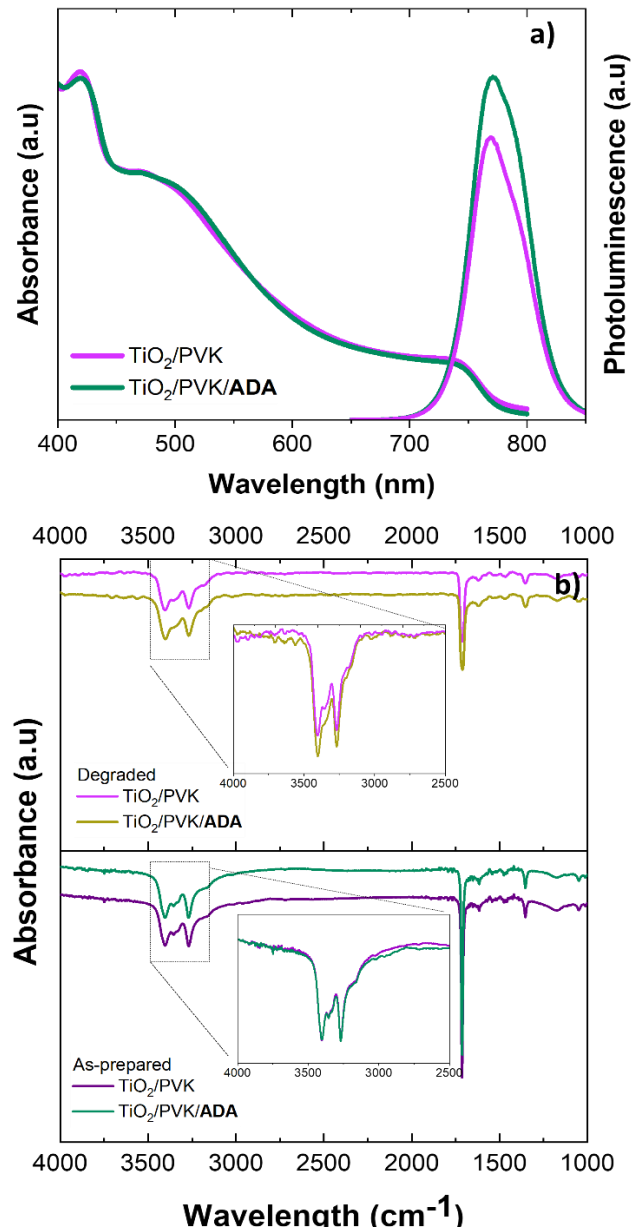

**Figure 2.** (a) Absorption and photoluminescence spectra before and after incorporating 6nm of adamantane-based ultrathin plasma polymer. (b) FTIR spectra of ADA-passivated perovskite films and reference samples before and after exposure to 77% relative humidity for 15 days.

Additionally, due to the partially hydrophobic nature of the ADA layer, the protective effect of this passivating layer against ambient humidity was also evaluated. To assess this, the evolution of the FTIR spectra of the samples exposed to 77% relative humidity at room temperature over a period of 15 days was monitored (see Figure 2b). No significant changes are observed in the FTIR analysis for the fresh ADA samples. This indicates that the primary vibrational modes of the perovskite layer within the 4000-600 cm$^{-1}$ range, mainly associated with the bending and rocking modes, remain unaltered with the incorporation of the ADA layer. Furthermore, the peak intensities are comparable in both cases. However, it is observed that the N-H vibrational peaks (at 3406 and 3270 cm$^{-1}$) associated with -NH$_3^+$ decreased in the TiO$_2$/PVK sample,[47] while they remain stable in the TiO$_2$/PVK/ADA after exposure to the humidity test. The drop in peak intensity is attributed to the conversion of ammonium ions (NR$_4^{3+}$) to amines group (NR$_3$) within the organic cations of the perovskite materials. This decrease in N-H content serves as an indicator of the degradation of the perovskite material.[9,48,49] The degradation of perovskite films after the humidity testing was also assessed using XRD. Figure S3 shows the XRD patterns, where the characteristic diffraction peaks of the tetragonal perovskite phase (14.1°, 28.5°, and 31.9°) are observed in both samples.[5,50] Additional diffraction peaks corresponding to the formation of a hydrated perovskite complex (10.5°) and PbI$_2$ (12.7°) were also observed, with greater intensity in the TiO$_2$/PVK sample.[5,51]

Based on the above, the stability of the solar devices incorporating the ADA layer at the perovskite/spiro-OMeTAD interface was assessed under high humidity environments (70% relative humidity at room temperature) and continuous light exposure (ISOS-L-1). Figure S4 depicts the evolution of the average PV parameters after 100 hours under these high humidity conditions for two PSCs of each configuration. It is important to emphasize that the PSCs were tested without encapsulation. In this context, the ADA layer at perovskite/HTL interface was found to significantly enhance the stability in humid environments. Specifically, the device based on 6 nm ADA interlayer preserved 60% of the initial efficiency, whereas the efficiency of reference sample decreased to below 40%. The *FF* and $V_{OC}$ retained more than 90% of their initial values for 6 nm ADA samples, compared to a drop of 40% and 20%, respectively, in the reference samples. The trend for photocurrent is similar in both configurations, these were reduced by around 40%. The enhanced stability of ADA samples confirms our previous results demonstrating the protective nature of the plasma polymer layer against humidity-induced degradation, as shown in Figure 2b. [28,34]

In order to further analyze the mechanism by which the ADA protection works, impedance spectroscopy measurements were carried out. Results for impedance spectra obtained under white and monochromatic LED illumination and quasi open-circuit conditions are presented in **Figure 3**.

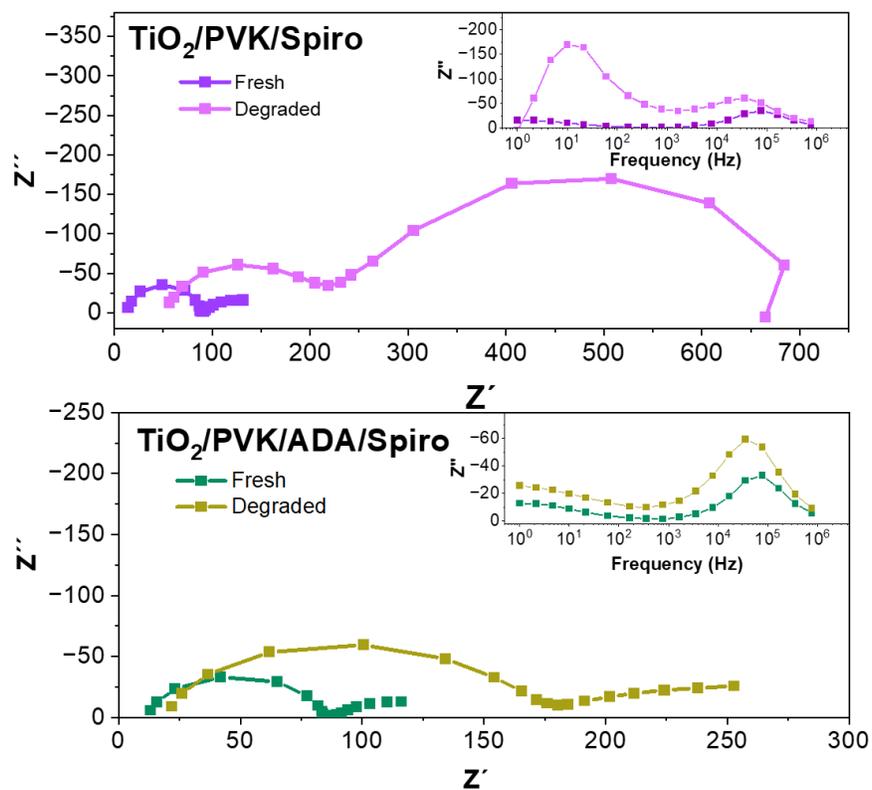

**Figure 3.** Impedance spectra obtained under LED illumination and quasi open circuit conditions for reference (TiO$_2$/PVK/Spiro/Au) and ADA-treated (TiO$_2$/PVK/ADA/Spiro/Au) perovskite cells. Results for fresh and degraded cells obtained using the same white LED light intensities are shown. Degradation was induced by storing the cells at ambient conditions for more than 1100 hours. Impedance spectra are shown as imaginary plots (Nyquist) and Balso as frequency plots in the insets.

The results show a dramatic change of the spectra shape for the reference samples, especially in the low frequency region (low frequency peaks and arcs) upon degradation. In contrast, the impedance response of the ADA samples proved to be much more robust in this respect. As it is in the low frequency region where ion-mediated and interfacial recombination phenomena are revealed, [52–55] the results suggest that ADA layer contributes to the stabilization of the interface by preventing the formation of mobile ionic defects. This is evidenced by the dramatic enlargement of the low frequency arc and the shift of the low frequency peak towards higher frequencies observed in the unprotected samples. A very similar effect in the impedance spectra was reported before, where ionic defects were intentionally introduced in the perovskite layer.[22]

The formation of ionic mobile charge upon degradation also explains the deterioration of the cell in terms of recombination, which is the reason behind the drop in the open circuit voltage. This is further confirmed by the analysis of the high-frequency parameters extracted from the impedance (see Figure S5). The high frequency resistance, as obtained from fits of the spectra to a simple R-(RC)-(RC) equivalent circuit (inset Figure S5),[46] significantly decrease for the degraded samples with respect to the fresh solar cells. This decrease is less pronounced for the cells that include ADA in the HTL interface. As previously reported, the formation and accumulation of mobile ions at interfaces leads to an enhancement of bulk and surface recombination routes.[22,24,56,57] The creation of new recombination mechanisms is also evidenced by the change of slope of the recombination resistance, quantified by the so-called apparent ideality factor as described by Bennet et al.[58] This parameter provides information about the recombination mechanism but is also affected when a large concentration of ions is present. In this work, apparent ideality factor values between 1.5 and 2 were obtained for the fresh samples for both the reference and ADA samples, indicative of bulk recombination as dominant mechanism but with a significant contribution of surface recombination losses. This is in line with the relative coincidence of results obtained with different LEDs (see Figure S5).[59] Interestingly, degradation of the reference samples led to apparent ideality factor values between 10 and 11 (see Table S3), pointing to a dramatic interference of ion accumulation at interfaces, as mentioned earlier. In striking contrast with these results, ADA samples kept similar values of the apparent ideality factor for degraded samples.

## Adamantane Plasma Polymers Layers for Both ETL/Perovskite and Perovskite/HTL Interfaces

In a previous work, we demonstrated that this ultrathin plasma polymer works as an effective passivation interface between the ETL and perovskite, enhancing both the moisture stability and reproducibility of solar devices.[34] Building on this, MHPSCs were prepared by sandwiching the perovskite layer between the optimal ADA layers for both interfaces (ETL/perovskite and perovskite/HTL) (see Scheme 1), and their stability in a humid environment was compared to reference samples (without any ADA passivating layer). On the one hand, **Figure 4a** shows the *J-V* curve of the most efficient perovskite solar cell with the ADA-based sandwiched architecture, as well as that of a reference sample. Table S4 and Figure S6 compare the PV parameters of six perovskite solar cells of each configuration. It was observed that the PV performance is not significantly affected by the double incorporation of the ADA passivation interlayers. On the other hand, Figure 4b illustrates the evolution of the PV parameters of samples subjected to an accelerated stability assessment, which was conducted under 100% relative humidity at room temperature and continuous illumination of 1 sun AM 1.5 G (ISOS-L-1). It was observed that ADA-double passivated samples retained nearly 80% efficiency of their initial efficiency after 4000 min in the stability test, whereas the reference sample decreased to 30% efficiency. It is important to remark that these results correspond to samples without encapsulation. Analyzing the evolution of the PV parameters, the efficiency drop observed in the ADA-based sandwiched samples was mainly caused by the loss of *FF* during the stability test, as both $V_{OC}$ and $J_{SC}$ retained their initial values. In contrast, although the reference sample maintained its initial $V_{OC}$, the $J_{SC}$ dramatically decreased to 50%. Comparing the stability results of ADA devices passivated only at the perovskite/HTL interface (Figure S4) with those of ADA-double passivated samples (Figure 4), the latter demonstrated better resistance to humid environments, mainly due to their ability to maintain a constant photocurrent during the stability test. In this regard, it is well known that $TiO_2$ exhibits a catalytic effect when exposed to UV light, leading to perovskite decomposition.[14,60,61] Thus, the ADA double passivation strategy not only protected against humidity-induced degradation, but the ADA layer at ETL/perovskite interface could also mitigate the degradation pathway caused by the photocatalytic effect of TiO$_2$. To demonstrate the passivation effect of ADA layer on the UV light-assisted decomposition of perovskite, TiO$_2$/perovskite samples with and without ADA at TiO$_2$/perovskite interface were prepared and exposed to constant illumination of UV light at 375 nm, to ensure the TiO$_2$ activation. The photoluminescence (PL) emission at 765 nm was monitored over time, and the results are shown in Figure 4c. Figure S7 displays the steady-state PL spectra before and after stability test. It was observed that the sample containing ADA at TiO$_2$/perovskite interface retained the intensity of the emission peak after 1200 min of UV light exposure, while the reference sample lost all intensity within the same time.

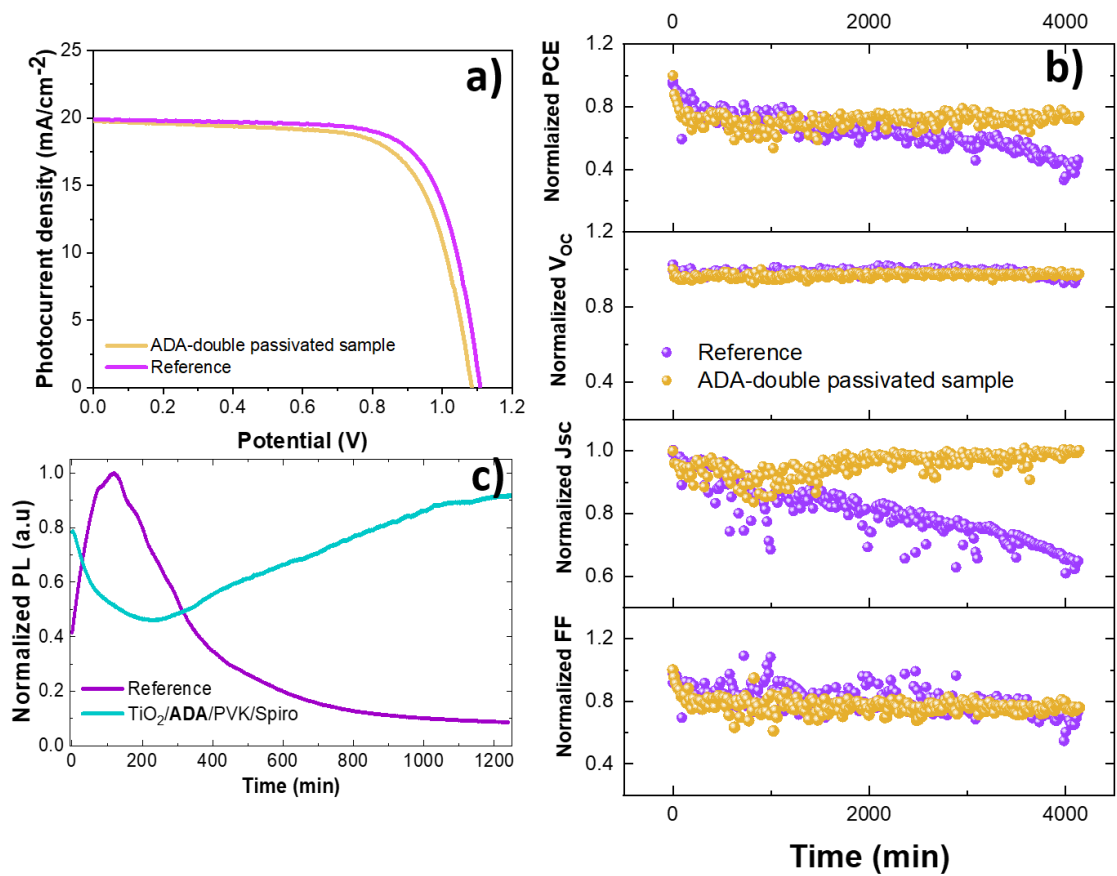

**Figure 4.** (a) JV curves measured under constant 1 sun AM 1.5G illumination for a representative perovskite solar cell with the ADA-based sandwiched architecture (TiO$_2$/ADA/PVK/ADA) and for a representative reference sample. The measurements were obtained at reverse scan using a 0.14 cm$^2$ mask. (b) Normalized PV parameters as a function of time obtained from the current-voltage curves. The measurements were carried out at 100

% relative humidity at room temperature of temperature. (c) Normalized photoluminescence emission peak at 765 nm as a function of time when the samples are exposed to 375 nm UV light.

## Conclusions

Adamantane plasma-polymeric films deposited by RPAVD technique have been evaluated as passivation layer at the perovskite/HTL interface in conventional *n-i-p* mesoporous perovskite solar cells. It was found that the incorporation of a 6 nm ADA layer at the perovskite/spiro-OMeTAD interface does not significantly alter the photovoltaic parameters of the solar cells. The exposition to a remote plasma during the RPAVD process does not affect the optical properties of the perovskite material, as evidenced by UV-Vis and FTIR spectroscopy. Additionally, the morphology of perovskite material remains unchanged, as demonstrated by SEM.

It has been also shown that a 6 nm ADA layer is insufficient to fully prevent water-induced degradation of the perovskite material; however, it does provide a degree of protection in high-humidity environments. In fact, the ADA interlayer at perovskite/HTL interface improves the stability of the perovskite solar cells under moist conditions (70% relative humidity at room temperature), retaining 60% of their initial efficiency, whereas the reference samples drop to 40% of their initial efficiency. The impedance analysis shows that this stability enhancement appears to be associated with the prevention of the formation of mobile ionic defects (passivation), which accumulate at interfaces and trigger additional recombination routes that, mainly, reduced the open circuit voltage for the unprotected reference solar cells. This finding aligns with the observed increase in steady-state photoluminescence intensity for the passivated samples.

On the other hand, protecting both the ETL/perovskite and perovskite/HTL interfaces with ultrathin ADA films results in solar devices that maintain nearly 80% of their initial efficiency after 4000 minutes of humidity testing (100% relative humidity at room temperature), whereas the reference sample dropped to 30% efficiency. The enhanced stability of the dual-passivated perovskite solar cell is not only due to the protective effect against humid environments, but also to the mitigation of degradation caused by the photocatalytic effect of $TiO_2$ under UV radiation.

This demonstrates the potential of ADA layers deposited via RPAVD technique as a versatile material, capable of acting both as a moisture barrier at interfaces and as a nanometric coating to improve the stability of perovskite solar cells. The ability of a single material to perform multiple functions within the device presents a practical advantage, which could facilitate the development of industrially scalable perovskite solar cells.

**Notes**

The authors declare no competing financial interest.

**Acknowledgments**

We thank the projects PID2022-143120OB-I00 and TED2021-130916B - I00, funded by MCIN/AEI/10.13039/501100011033 and by "ERDF (FEDER) a way of making Europe, the Fondos NextgenerationEU and Plan de Recuperación, Transformación y Resiliencia.", the M-ERA.NET 3 Program (Project Angstrom EU H2020 grant agreement No 958174) and the EU H2020 program under grant agreement 851929 (ERC Starting Grant 3DScavengers).

M.N, G. O and J.A.A. acknowledge the Ministerio de Ciencia e Innovación of Spain, Agencia Estatal de Investigación (AEI), EU (FEDER) under grants PID2022-140061OB-I00 (DEEPMATSOLAR) and CNS2022-135694 (IMPRESOL) and partial funding through the European Union's Horizon Europe research innovation program under the Platform-Zero project (grant agreement number: 101058459).

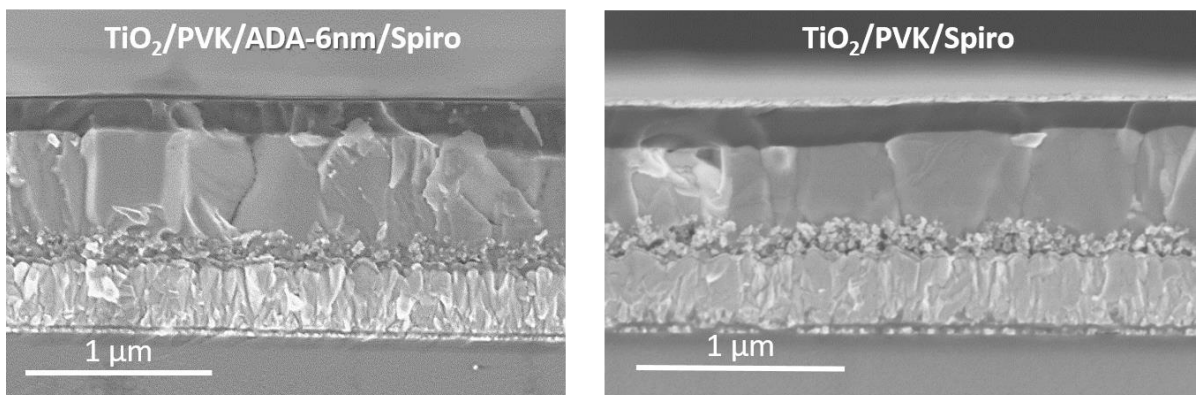

**Supporting information**

**Figure S1**. Cross-sectional SEM micrographs of complete perovskite solar cells with ADA polymer at the perovskite/HTL interface and reference sample (without ADA passivation layer).

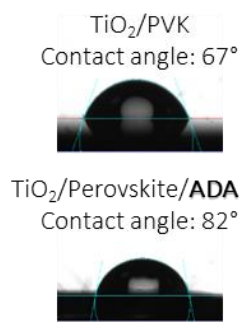

**Figure S2.** Static water contact angle before and after the treatment of ADA-6nm of m-TiO$_2$/PVK electrode for a milliQ water droplet of 1 µL.

**Table S1.** Photovoltaic statistical data obtained from *J-V* curves of perovskite solar cells incorporating ADA films of different thickness at the perovskite/HTL interface, measured during the reverse scan at 100 mW·cm$^{-1}$ under 1 sun - AM 1.5G illumination. In parentheses the photovoltaic data of the solar cells with maximum power conversion efficiency.

| Device | PCE (%) | FF (%) | $V_{OC}$ (V) | $J_{SC}$ (mA/cm$^2$) |
|---|---|---|---|---|
| ADA-30nm | 11.2 ±3.4 (13.1) | 64.7 ±12.9 (70.9) | 0.96 ±0.06 (0.99) | 17.6 ±1.5 (18.7) |
| ADA-20nm | 15.5 ±0.3 (15.8) | 72.1 ±1.1 (73.5) | 1.12 ±0.00 (1.11) | 19.3 ±0.2 (19.3) |
| ADA-6nm | 16.6 ±0.3 (17.4) | 73.2 ±1.2 (73.4) | 1.09 ±0.02 (1.13) | 20.7 ±0.3 (21.0) |
| Reference | 16.6 ±0.5 (17.4) | 74.1 ±1.4 (74.8) | 1.08 ±0.02 (1.10) | 20.7 ±0.4 (21.0) |

**Table S2.** Variable angle spectroscopic ellipsometry (VASE)

| Sample | Thickness | An | Bn | Cn |
|---|---|---|---|---|
| ADA-30nm | 31.2 | 1.60 | 0.04 | 0 |
| ADA-20nm | 23.7 | 1.63 | 0 | 0 |
| ADA-6nm | 6 | 1.65 | 0.06 | 0 |

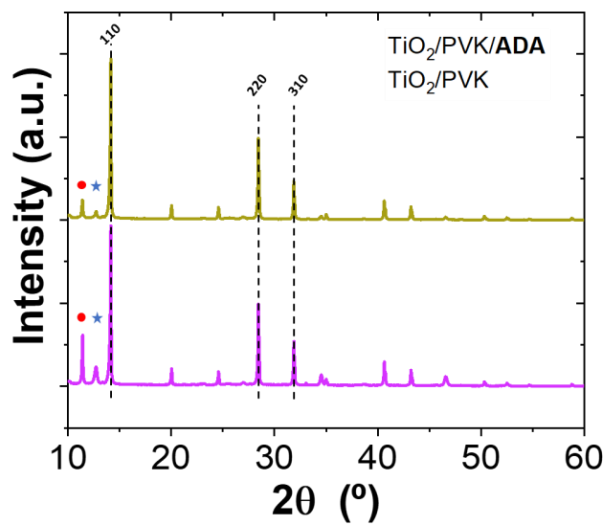

**Figure S3**. X-ray diffraction spectra of ADA-passivated perovskite films (TiO$_2$/PVK/ADA) and reference samples (TiO$_2$/PVK) before and after exposure to 77% relative humidity for 15 days. The positions of the XRD peaks for (blue star) PbI$_2$ and (red circle) the hydrated complex are marked in the graphs.

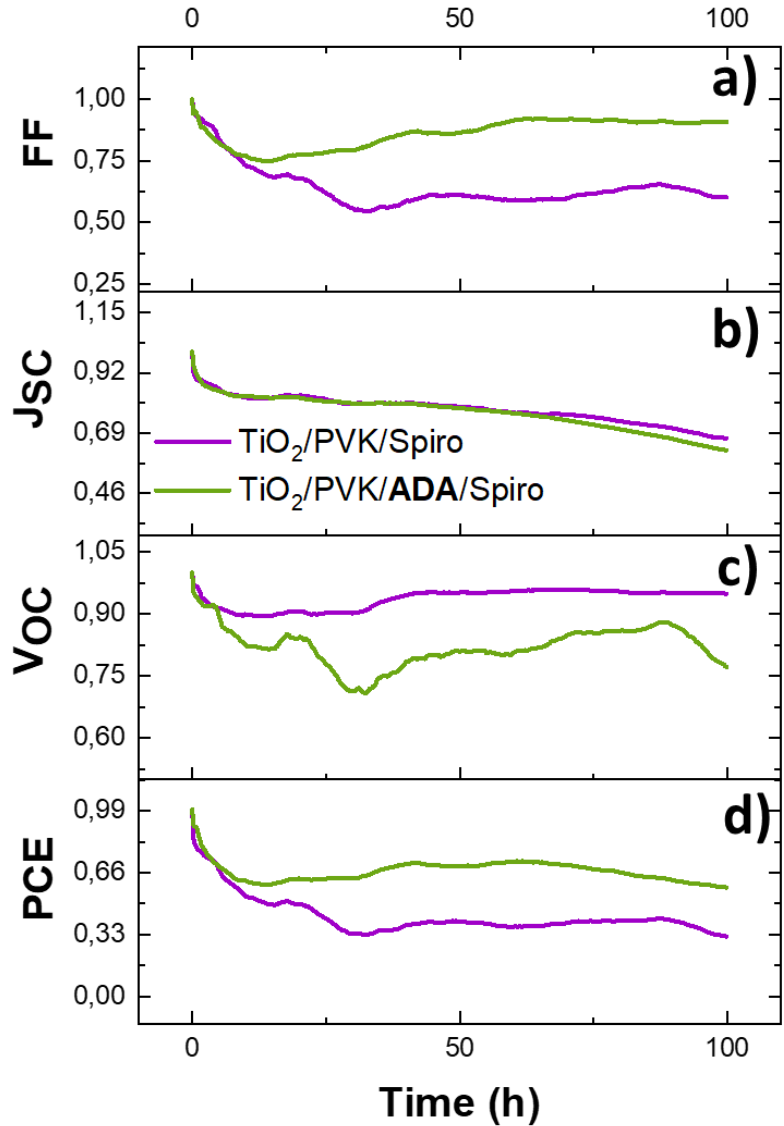

**Figure S4**. Normalized photovoltaic parameters as a function of time, obtained from average current-voltage curves for two solar cells of each type measured under constant 1 sun AM 1.5G illumination. The measurements were carried out at 80% relative humidity at room temperature. (a) fill factor (*FF*), (b) short-circuit current density ($J_{SC}$), (c) open-circuit voltage ($V_{OC}$) and (d) power conversion efficiency (*PCE*).

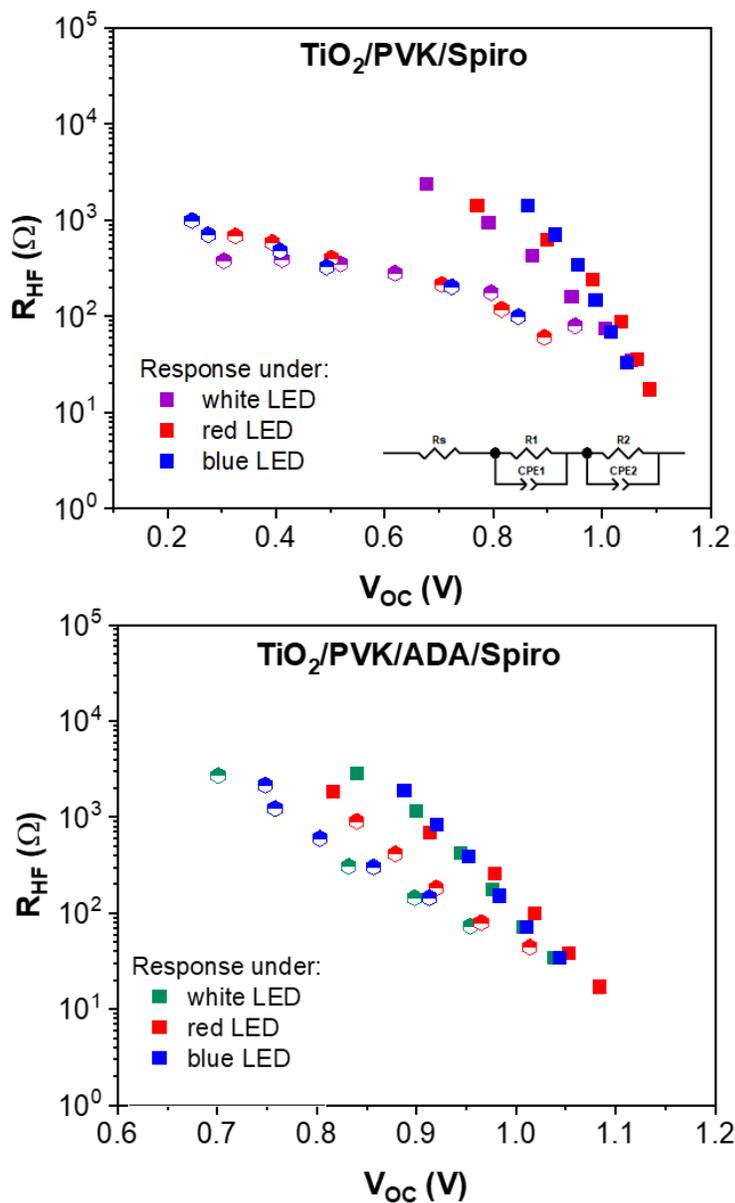

**Figure S5.** High frequency resistance values as obtained from the equivalent circuit (inset) fittings of the spectra shown in Figure 3 using three excitation wavelengths of blue LED (λ = 465 nm), red LED (λ = 635 nm) and white LED for reference (top) and ADA treated cells (bottom). The squares correspond to the fresh solar cells and the half-empty diamonds to degraded solar devices.

**Table S3**. Apparent ideality factors determined from the slope of the high-frequency resistance with respect to the open-circuit voltage for devices with a 6 nm ADA interlayer at the perovskite/Spiro-OMeTAD interface and reference devices. Measurements were conducted under red and blue illumination, both before and after the induced degradation process.

| Device | LED | Before | After |
|---|---|---|---|
| Reference | Red | 2.05 | 9.55 |
|  | Blue | 1.84 | 11.2 |
| TiO$_2$/PVK/ADA/Spiro | Red | 1.75 | 1.98 |
|  | Blue | 1.47 | 2.82 |

**Table S4.** Photovoltaic statistical data obtained from *J-V* curves of perovskite solar cells cell with the ADA-based sandwiched architecture and reference samples, measured during the reverse scan at 100 mW·cm$^{-1}$ under 1 sun - AM 1.5G illumination. In parentheses the photovoltaic data of the solar cells with maximum power conversion efficiency.

| Device | $V_{OC}$ (V) | $J_{SC}$ (mA/cm$^2$) | FF(%) | PCE (%) |
|---|---|---|---|---|
| Reference | 1.10 ± 0.03 | 19.2 ± 0.8 | 72 ± 3 | 15.2 ± 1.15 |
|  | (1.11) | (20.3) | (71) | (16.1) |
| ADA based sandwiched | 1.06 ± 0.03 | 19.0 ± 0.6 | 72 ± 3 | 14.3 ± 0.4 |
|  | (1.09) | (19.8) | (70) | (15) |

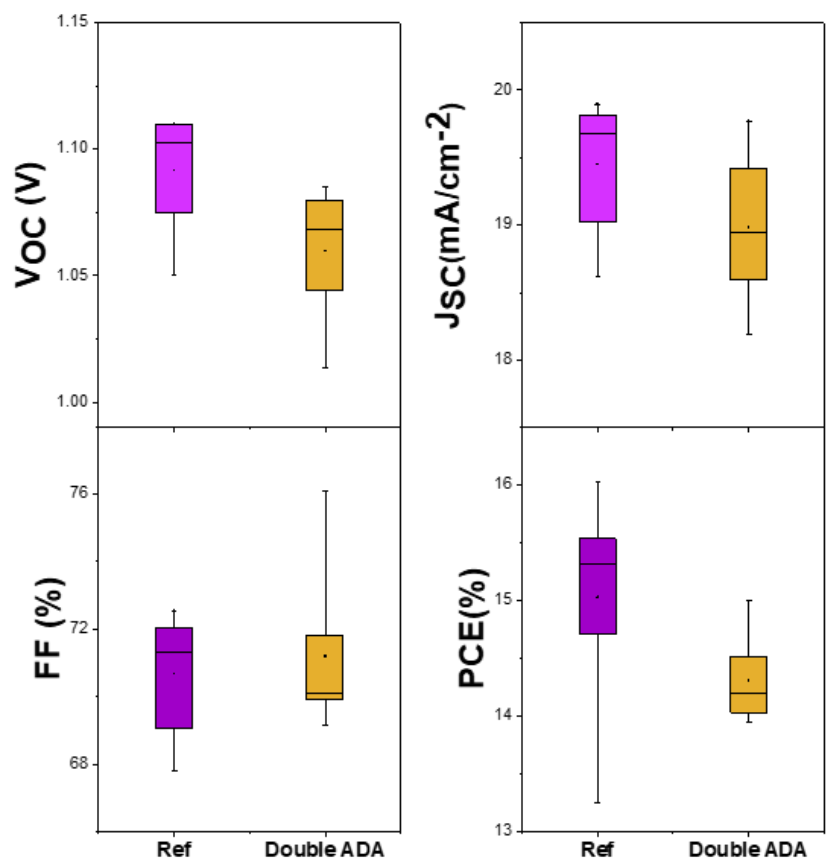

**Figure S6**. Statistics of the photovoltaic parameters by analysis of variance for reference samples and ADA double-passivated samples (TiO$_2$/ADA/PVK/ADA). These data have been obtained under sun AM 1.5G illumination, at 100 mV·s$^{-1}$ during the reverse scan and using a mask of 0.14 cm$^2$.

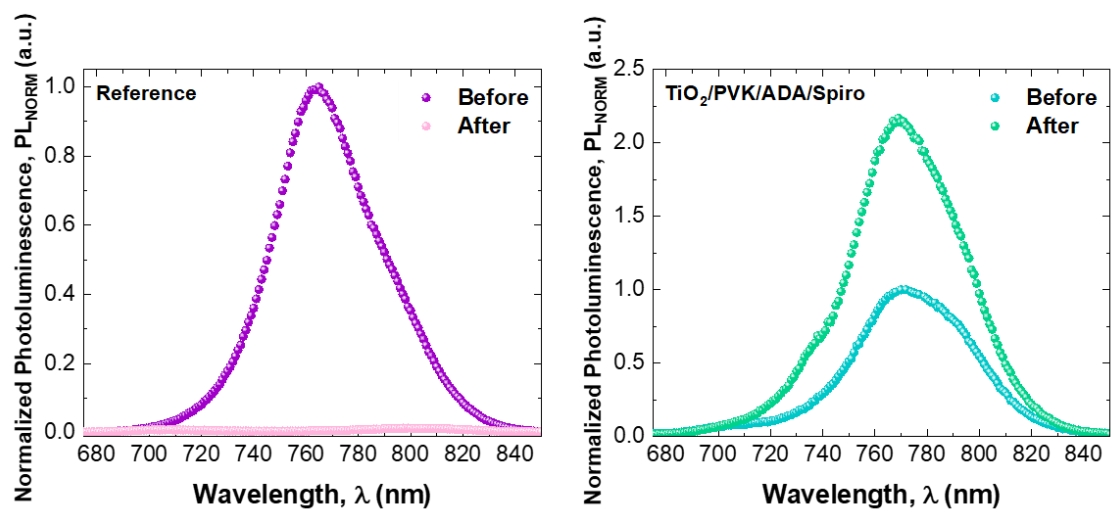

**Figure S7**. Steady-state PL spectra before and after the UV light-assisted decomposition test for the reference and ADA-based sample.